\documentclass[12pt,epsfig]{article}
\usepackage{epsfig}

\begin{document}

\begin{center}
{\Large {\bf The Masses of Gauge Bosons in $SU(3)_{C}\times SU(2)_{L}\times U(1)_{Y^{\prime}}\times U(1)_{B-L}$ gauge model.}}

M. C. Rodriguez \\
{\it Grupo de F\'\i sica Te\'orica e Matem\'atica F\'\i sica \\
Departamento de F\'\i sica \\
Universidade Federal Rural do Rio de Janeiro - UFRRJ \\
BR 465 Km 7, 23890-000, Serop\'edica, RJ, Brazil, \\
email: marcoscrodriguez@ufrrj.br} 
\end{center}

\begin{abstract}
We will present within the context of the supersymmetric models with  
$SU(3)_{C}\times SU(2)_{L}\times U(1)_{Y^\prime}\times U(1)_{B-L}$ gauge symmetry an explanation for the new data on the $W$-boson mass 
recently presented by the CDF collaboration. We will also study the 
neutral boson sector of this model.
\end{abstract}

Keywords: Supersymmetric models, Extensions of Electroweak Gauge sector, 
Non-standard-model neutrinos, right-handed neutrinos, etc. \\
PACS number(s): 12.60. Jv, 12.60.Cn, 14.60.St.

\section{Introduction}
\label{sec:intro}

In the the Standard Model (SM), the gauge symmetry is defined 
as \cite{sg,Kronfeld:2010bx}
\begin{equation}
SU(3)_{C}\times SU(2)_{L}\times U(1)_{Y}
\end{equation}
and the gauge sector of this model is presented in our 
Tab.(\ref{tablegaugebosonssm}).
\begin{table}[h]
\begin{center}
\begin{tabular}{|c|c|c|c|c|}
\hline
${\rm{Group}}$ & ${\rm{Gauge \,\ Bosons}}$ & ${\rm Gauge \,\ constant}$ \\
\hline
$SU(3)_{C}$ & $g^{a}_{m}$ & $g_{s}$  \\
\hline
$SU(2)_{L}$ & $W^{i}_{m}$ & $g$ \\
\hline
$U(1)_{Y}$ & $b_{m}$ & $g_{Y}$ \\
\hline
\end{tabular}
\end{center}
\caption{Information on fields contents of each vector superfield of this model, where $a=1,2, \ldots ,8$ and $i=1,2,3$.}
\label{tablegaugebosonssm}
\end{table}

There are two massives gauge boson\footnote{See our Eq.(\ref{tablemssmrpv}) in 
App.(\ref{sec:mssmrpv})}. The charged gauge boson $W^{\pm}$
\begin{eqnarray}
M_{W}= \frac{gv}{2}, 
\label{wmasssm}
\end{eqnarray}
theoretical estimates indicate the following numerical value for its mass  
\begin{equation}
\left( M_{W} \right)_{\rm SM}= \left(
80.3505 \pm 0.0077 \right) \,\ {\mbox GeV.}
\label{smvalue}
\end{equation}
however the average value of experimental 
measurements provides us with the following data \cite{average} 
\begin{eqnarray}
\left( M_{W} \right)_{\rm EXP}&=& \left(
80.4133 \pm 0.0080 \right) \,\ {\mbox GeV}. 
\label{average}
\end{eqnarray}

Recently, Fermilab's CDF collaboration presented its highly 
precise measurements, with an accuracy of the order of 
$\sim 10^{-4}$, for the mass of the $W$-boson mass \cite{cdf}
\begin{equation}
\left( M_{W} \right)_{\rm CDF}= \left(
80.4335 \pm 0.0094 \right) \,\ {\mbox GeV,}
\label{cdfresult}
\end{equation}
this value represents an increase greater than $6$ $\sigma$ 
in relation to the value predicted by the SM. It is clear that, if this 
result is proven by other experimental collaborations, we have a 
strong indication of physics beyond SM \cite{Rodriguez:2022hsj}.

Recently it was proposed this shift can be easily explained by a real 
triplet Higgs 
boson \cite{Blank:1997qa,Chen:2006pb} and with  
an extra complex triplet Higgs boson \cite{evans} and also at leptoquark 
model \cite{Bhaskar:2022vgk}. 
There is an attempt to explain this anomaly in the context of the 
Minimal $R$-Symmetric extension of MSSM, known as 
MRSSM \cite{Kribs:2007ac,Diessner:2014ksa,Diessner:2019ebm,athron} and also, the use of Two Higgs Doulet Model (2HDM) \cite{Lu:2022bgw}, and 
there is also a top-down motivated model in which extra sector states from a D3-brane also 
accommodate this result \cite{Heckman:2022the}. 

Two  important problems  of today's particle physics are : explain the 
masses of neutrinos as well as generate dark matter. 
Today we know that neutrinos have mass, that  their flavors are mixed and also 
that there is at least one Higgs boson. However, we do not know yet if 
neutrinos are Majorana or Dirac particles and if there are more neutral 
scalars. Moreover, right-handed sterile neutrinos i.e., singlets under the 
SM, have not been observed yet. However, on the other hand, there is 
strong experimental evidence for the existence of dark matter and it cannot 
be resolved within the SM.  

Certainly the most popular extension of the SM is its supersymmetric 
counterpart called Minimal Supersymmetric Standard Model (MSSM)
\cite{Rodriguez:2019mwf,kazakov1}. The main 
motivation to study this model is that it provides a solution to the hierarchy 
problem by protecting the electroweak scale from large radiative 
corrections.

Supersymmetric theories (SUSY) has also made several correct predictions 
\cite{Chung:2003fi}
\begin{itemize}
\item SUSY predicted in the early 1980s that the top quark would be heavy;
\item SUSY GUT theories with a high fundamental scale accurately predicted the present experimental value of $\sin^{2} \theta_{W}$ 
before it was mesured;
\item SUSY requires a light Higgs boson to exist.
\end{itemize}
Together these success provide powerful indirect evidence that low energy SUSY is indeed part of correct description of nature. The current status of 
the search for supersymmetry is presented in reference \cite{gladkaza}.

In this model, as in the SM, neutrinos are 
massless, but in this model there are interactions that volates Lepton 
or Baryon Number conservation, which allows us to generate non-zero masses 
for at least one of the neutrinos as well as have an explanation for 
the Pontecorvo-Maki-Nakagawa-Sakata (PMNS) matrix is also 
accommodated \cite{Rodriguez:2022gfq}.

Some years ago, it was proposed  a model in which 
$U(1)_{(B-L)}$ is not just a new factor added to the 
SM gauge symmetry but $U(1)_{Y}$ is 
substituted by $U(1)_{Y^{\prime}}\times U(1)_{(B-L)}$  and 
the breaking 
\begin{equation}
U(1)_{Y^{\prime}}\times U(1)_{(B-L)}\to U(1)_{Y}
\end{equation} 
occurs at the TeV scale. We choose the $Y^{\prime}$ parameter to obtain the 
hypercharg $Y$ of the SM through the following expression
\begin{equation}
Y=Y^{\prime}+(B-L).
\end{equation} 

Moreover, the number of right-
handed neutrinos, $N_{iR}$, and their $(B-L)$ (or 
$Y^{\prime}$) quantum numbers are free parameters,  but the  
cancellation of the cubic and the linear anomalies implies 
that at least  three right-handed neutrinos must be added 
to the matter representation content. Explicit solutions 
for the $(B-L)$ (or $Y^{\prime}$) parameters show that at 
least two types of model arise~\cite{Montero:2007cd} 
\begin{itemize}
\item[1-)] the model with three 
right-handed neutrinos having  $(B-L)=-1$ (known as three 
identical neutrinos); 
\item[2-)] the model with two right-handed 
neutrinos having $(B-L)=-4$ and the third one having $(B-L)=5$ 
(known as three non-identical neutrinos). 
\end{itemize}

The Supersymmetric $U(1)_{Y^{\prime}}\times U(1)_{B-L}$ model 
with three identical neutrinos was presented 
at \cite{Montero:2016qpx}, where it was considered 
that the sneutrinos\footnote{They are the usual superpartner of 
neutrinos introduced in SM.}, both left-handed, $\tilde{\nu}_{iL}$, and 
right-handed, $\tilde{N}_{iR}$, have vaccum expectation values (VEV) equal to zero. The supersymmetric 
model with nonidentical neutrinos is in Ref~\cite{Montero:2016qpxno}. In 
both supersymmetric models we can explain the masses of neutrino as 
well we have interesting candidate for dark model 
\cite{Montero:2016qpx,Montero:2016qpxno}. There is an 
interesting model having fractional $(B-L)$ charges \cite{Bernal:2018aon}.

In this article 
we will relax this hypothesis and show that when we allow our left-handed 
sneutrinos get VEV, 
and let's represent it as $v^{L}_{3}$, we can explain the new 
CDF data about $W$-boson mass. We will study here the masses of 
Gauge Bososn in both supersymmetric models mentioned above.

The outline of this paper is as follows. In Sec.\ref{sec:identical} we study the masses of the gauge bosons in the model with three identical 
right-handed neutrinos and then we perform the same analyses for the case with three noidentical right-handed neutrinos. Our
conclusions appear in Sec.~\ref{sec:con}. We will briefly present the MSSMRV to 
better understand the results of charged bosons and show a previous analysis of 
the neutral boson sector to better understand the results of our models in 
App.(\ref{sec:mssmrpv}). In the last Appendix we present 
numerical analysis to understand the mixing in the sector of 
neutral gauge bosons for three non-identical neutrinos, the 
App.(\ref{partialresultsnonidentical}).

\section{Model with three identical neutrinos.}
\label{sec:identical}
The gauge symmetry of the model is given by
\begin{equation} 
SU(2)_{L}\times U(1)_{Y^{\prime}}\times U(1)_{(B-L)},
\end{equation}
the charge operator is defined using the following algebraic 
expression~\cite{Montero:2007cd}
\begin{equation}
\frac{Q}{e}=I_{3}+ \frac{1}{2} \left[ Y^{\prime}+(B-L) \right] .
\end{equation}
We also assume that the $(B-L)$ and $Y^{\prime}$ assignments are restricted 
to  integer numbers, as we mentioned in our introduction.

We introduce the charged leptons
\footnote{We are omitting the quarks, if you are interested in them, see~\cite{Montero:2007cd}.} as well as the usual MSSM scalars in the 
following chiral superfields
\begin{eqnarray}
\hat{L}_{i} &=& \left(
\begin{array}{c}
\hat{\nu}_{i} \\
\hat{l}_{i}
\end{array}
\right)_{L} \sim ({\bf 1},{\bf 2},0,-1), \,\
\hat{E}_{iL} \sim ({\bf 1},{\bf 1},1,1), \,\ i=1,2,3, \nonumber \\
\hat{H}_{1} &=& \left(
\begin{array}{c}
\hat{h}^{+}_{1} \\
\hat{h}^{0}_{1}
\end{array}
\right) \sim ({\bf 1},{\bf 2},+1,0), \,\
\hat{H}_{2} = \left(
\begin{array}{c}
\hat{h}^{0}_{2} \\
\hat{h}^{-}_{2}
\end{array}
\right) \sim ({\bf 1},{\bf 2},-1,0), \nonumber \\
\label{leptonsm1}
\end{eqnarray}
with $i=1,2,3$, and in parentheses we present the transformations 
properties under the respective gauge factors  
$(SU(3)_{C},SU(2)_{L},U(1)_{Y^{\prime}},U(1)_{(B-L)})$. We use the standard notation $\hat{E}_{iL}\equiv (\hat{l}_{iR})^{c}$. In this model, we also 
introduce three right-handed neutrinos \cite{Montero:2016qpx}
\begin{equation}
\hat{N}_{iL} \sim ({\bf 1},{\bf 1},-1,1),
\label{righthandedneutrinosid}
\end{equation}
where again $\hat{N}_{iL}\equiv (\hat{\nu}_{iR})^{c}$ as well as two singlet $SU(2)_{L}$ scalars 
\begin{eqnarray}
\hat{\phi}_{1} &\sim& ({\bf 1},{\bf 1},-2,2), \,\
\hat{\phi}_{2} \sim ({\bf 1},{\bf 1},2,-2).
\label{esc1model}
\end{eqnarray}

The  particle content of each chiral superfields defined above are 
presented in the Tab.(\ref{lepm1}), we also show their quantum numbers.
\begin{table}[h]
\begin{center}
\begin{tabular}{|c|c|c|}
\hline
$\mbox{ Chiral Superfield} $ &  $\mbox{ Scalars} $ & $\mbox{ Fermions} $ 
\\ \hline
$\hat{L}_{i}\sim ({\bf 1},{\bf 2},0,-1)$ & 
$\tilde{L}_{i}\sim ({\bf 1},{\bf 2},0,-1)$ & 
$L_{i}\sim ({\bf 1},{\bf 2},0,-1)$ \\ \hline
$\hat{E}_{i}\sim ({\bf 1},{\bf 1},1,1)$ & 
$\tilde{E}_{i}\sim ({\bf 1},{\bf 1},1,1)$ & 
$E_{i}\sim ({\bf 1},{\bf 1},1,1)$  \\ \hline
$\hat{N}_{i}\sim ({\bf 1},{\bf 1},-1,1)$ & 
$\tilde{N}_{i}\sim ({\bf 1},{\bf 1},-1,1)$ & 
$N_{i}\sim ({\bf 1},{\bf 1},-1,1)$ \\ \hline
$\hat{H}_{1}\sim ({\bf 1},{\bf 2},+1,0)$ & 
$H_{1}\sim ({\bf 1},{\bf 2},+1,0)$ & 
$\tilde{H}_{1}\sim ({\bf 1},{\bf 2},+1,0)$ \\ \hline
$\hat{H}_{2}\sim ({\bf 1},{\bf 2},-1,0)$ & 
$H_{2}\sim ({\bf 1},{\bf 2},-1,0)$ & 
$\tilde{H}_{2}\sim ({\bf 1},{\bf 2},-1,0)$ \\ \hline
$\hat{\phi}_{1}\sim ({\bf 1},{\bf 1},-2,2)$ & 
$\phi_{1}\sim ({\bf 1},{\bf 1},-2,2)$ & 
$\tilde{\phi}_{1}\sim ({\bf 1},{\bf 1},-2,2)$ \\ \hline
$\hat{\phi}_{2}\sim ({\bf 1},{\bf 1},2,-2)$ & 
$\phi_{2}\sim ({\bf 1},{\bf 1},2,-2)$ & 
$\tilde{\phi}_{2}\sim ({\bf 1},{\bf 1},2,-2)$ \\ \hline
\end{tabular}
\end{center}
\caption{Particle content in each chiral superfield introduced in 
Eqs.(\ref{leptonsm1},\ref{righthandedneutrinosid},\ref{esc1model}).}
\label{lepm1}
\end{table}

The following neutral scalars in doublet representation get non-zero vaccum 
expectation value (VEV) denoted as
\begin{eqnarray}
\langle H_{1} \rangle &\equiv& \frac{1}{\sqrt{2}}\left[ v_{1}+\sigma^{0}_{1}+ \imath \varphi^{0}_{1} \right], \,\
\langle H_{2} \rangle \equiv \frac{1}{\sqrt{2}}\left[ v_{2}+\sigma^{0}_{2}+ \imath \varphi^{0}_{2} \right], \nonumber \\
\langle \tilde{L}_{3} \rangle &\equiv& \frac{1}{\sqrt{2}}\left[ v^{L}_{3}+ \tilde{\nu}^{R}_{3}+ \imath \tilde{\nu}^{I}_{3} \right], 
\label{vevyprcomum}
\end{eqnarray}
which will be constrained by the relation $(v_{1})^{2}+(v_{2})^{2}+(v^{L}_{3})^{2}=(246\, \textrm{GeV})^{2}$, see our 
Eq.(\ref{wmassidentical}) and Fig.(\ref{fgb2}). The scalars in 
singlet representation get the following VEV
\begin{equation}
\langle \phi_{1}\rangle = \frac{1}{\sqrt{2}}\left[ u_{1}+\sigma^{0}_{4}+ \imath \varphi^{0}_{4} \right], \,\
\langle \phi_{2}\rangle = \frac{1}{\sqrt{2}}\left[ u_{2}+\sigma^{0}_{5}+ \imath \varphi^{0}_{5} \right].
\label{soidentical}
\end{equation}
The values of $u_{1}$ and $u_{2}$ that give the scale of the $U(1)_{(B-L)}$ symmetry breaking are not fixed, they may have 
values ranging from TeV to much higher scales.

Concerning the gauge bosons and their superpartners, they are introduced in 
vector superfields, as we shown in Tab.(\ref{table2}), together with 
the gauge coupling constant of each group.
\begin{table}[h]
\begin{center}
\begin{tabular}{|c|c|c|c|c|}
\hline
${\rm{Group}}$ & ${\rm Vector \,\ Superfield}$ & ${\rm{Gauge \,\ Bosons}}$ & ${\rm{Gaugino}}$ & ${\rm Gauge \,\ constant}$ \\
\hline
$SU(3)_{C}$ & $\hat{G}^{a}$ & $g^{a}_{m}$ & $\tilde{g}^{a}$ & $g_{s}$  \\
\hline
$SU(2)_{L}$ & $\hat{W}^{i}$ & $W^{i}_{m}$ & $\tilde{W}^{i}$ & $g$ \\
\hline
$U(1)_{Y^{\prime}}$ & $\hat{b}_{Y^{\prime}}$ & $(b_{Y^{\prime}})_{m}$ & $\tilde{b}_{Y^{\prime}}$ & $g_{Y^{\prime}}$ \\
\hline
$U(1)_{(B-L)}$ & $\hat{b}_{BL}$ & $(b_{BL})_{m}$ & $\tilde{b}_{BL}$ & $g_{BL}$ \\
\hline
\end{tabular}
\end{center}
\caption{Information on fields contents of each vector superfield of this model.}
\label{table2}
\end{table}

\begin{figure}
\centering
\includegraphics[width=0.8\textwidth]{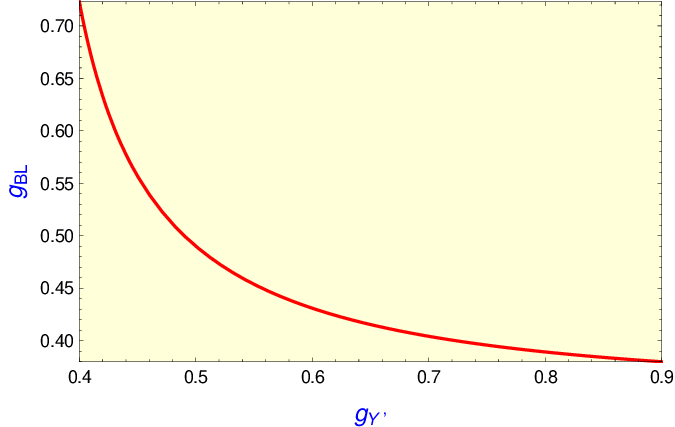}        
\caption{The values for $g_{BL}$ as function of $g_{Y^{\prime}}$, as 
defined in our Eq.(\ref{runcouplingconstant}).} 
\label{gblvsgyp}
\end{figure}

\begin{figure}
\centering
\includegraphics[width=0.8\textwidth]{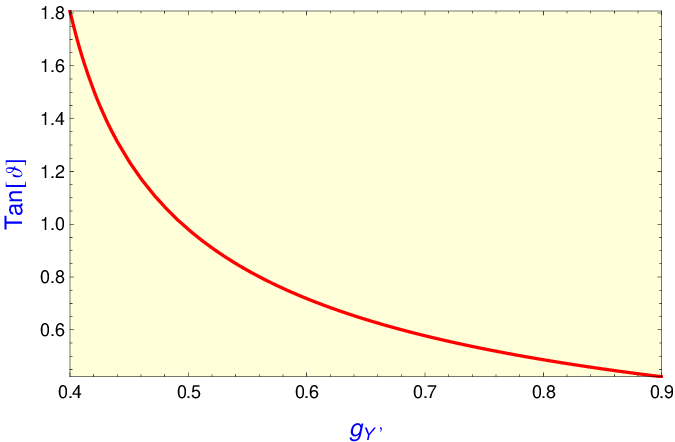}        
\caption{The values for $\tan \vartheta$ as function of $g_{Y^{\prime}}$, as 
defined in our Eq.(\ref{defalphaprime}).} 
\label{tHtvsgyp}
\end{figure}

\begin{figure}
\centering
\includegraphics[width=0.8\textwidth]{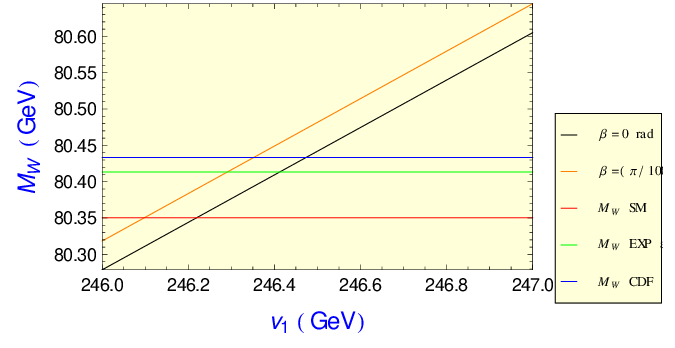}        
\caption{The masses of $W$ gauge boson as fuction of $v_{1}$ for 
$\theta = 0$ rad, and some values of our $\beta$-parameter shown on the box. The value of line red is given by Eq.(\ref{smvalue}), the green by Eq.(\ref{average}) and the blue line 
for Eq.(\ref{cdfresult}).} 
\label{fgb2}
\end{figure}

The  gauge coupling constants are related  by the
following relation
\begin{eqnarray}
\frac{1}{g^{2}_{Y}}= \frac{1}{g^{2}_{Y^{\prime}}}+ \frac{1}{g^{2}_{BL}},
\label{runcouplingconstant}
\end{eqnarray}
where $g_{Y}$ is the SM $U(1)_{Y}$ coupling constant, see Eq.(\ref{gySM}). 
We can also define the $\vartheta$-parameter in the following way
\begin{equation}
\tan \vartheta \equiv \frac{g_{BL}}{g_{Y^{\prime}}},
\label{defalphaprime}
\end{equation}
In our Figs.(\ref{gblvsgyp},\ref{tHtvsgyp}) and Tab.(\ref{tablegblvsgyp}), 
we show the possibles values for $g_{BL}$ and $\tan \vartheta$ as function 
of $g_{Y^{\prime}}$ and where we can see that both couplings have the same 
order of magnitude, that is, they satisfy
\begin{equation}
{\cal O}(g_{Y^{\prime}})\sim {\cal O}(g_{BL}),
\end{equation}
as showed in \cite{Sanchez-Vega:2014rka}. 

\begin{table}[h]
\begin{center}
\begin{tabular}{|c|c|c|}
\hline
$g_{Y^{\prime}}$ & $g_{BL}$ & $\tan \vartheta$ \\
\hline
0.4 & 0.72 & 1.807 \\
\hline
0.485 & 0.506 & 1.042 \\
\hline
0.5 & 0.49 & 0.980 \\
\hline
0.6 & 0.43 & 0.718 \\
\hline
0.7 & 0.40 & 0.577 \\
\hline
0.8 & 0.39 & 0.487 \\
\hline
0.9 & 0.38 & 0.422 \\
\hline
\end{tabular}
\end{center}
\caption{The values for $g_{BL}$ and $\tan \vartheta$ as function 
of $g_{Y^{\prime}}$, as defined in our 
Eqs.(\ref{runcouplingconstant},\ref{defalphaprime}).}
\label{tablegblvsgyp}
\end{table}

We will not write the Lagrangian of the model here, it can be seen 
in \cite{Montero:2016qpx}. We calculate the mass of gauge bosons using the following expression 
\begin{eqnarray}
&&({\cal D}_{m}H_{1})^{\dagger}({\cal D}^{m}H_{1})+ 
({\cal D}_{m}H_{2})^{\dagger}({\cal D}^{m}H_{2})+
({\cal D}_{m}\tilde{L}_{iL})^{\dagger}
({\cal D}^{m}\tilde{L}_{iL})\nonumber \\ &+&
({\cal D}_{m}\tilde{N}_{iL})^{\dagger}
({\cal D}^{m}\tilde{N}_{iL})+ 
({\cal D}_{m}\phi_{1})^{\dagger}({\cal D}^{m}\phi_{1})+
({\cal D}_{m}\phi_{2})^{\dagger}({\cal D}^{m}\phi_{2}).
\end{eqnarray}

We get the following expression to the charged gauge bosons masses 
\begin{eqnarray}
M^{2}_{W}&=& \frac{g^{2}}{4} \left[ \left( v_{1} \right)^{2}+ \left( v_{2} \right)^{2}+ \left( v^{L}_{3} \right)^{2} \right] =
\frac{g^{2}v_{1}^{2}}{4} \left[ 1+ \tan^{2} \beta + \tan^{2} \theta \right].
\label{wmassidentical}
\end{eqnarray}
We define the parameters $\beta$, $\theta$ in the following way
\begin{eqnarray}
\tan \beta &=& \frac{v_{2}}{v_{1}}, \,\
\tan \theta = \frac{v^{L}_{3}}{v_{1}}.
\label{betadef} 
\end{eqnarray}
The parameter $\beta$ is introduced in the MSSM with $R$-Parity 
conservation \cite{Rodriguez:2019mwf,kazakov1} and also in the MSSM with 
$R$-Parity violation \cite{Rodriguez:2022gfq} and even in this case
\footnote{See our Append.(\ref{sec:mssmrpv}).}, as in this model, 
we also introduce a new angle $\theta$. We can also rewrite from 
Eq.(\ref{wmassidentical}) the following relation
\begin{eqnarray}
\delta M_{W} &=& \frac{gv_{1}}{2}  \tan \theta .
\label{wmassdifidentical}
\end{eqnarray}

\begin{figure}
\centering
\includegraphics[width=0.8\textwidth]{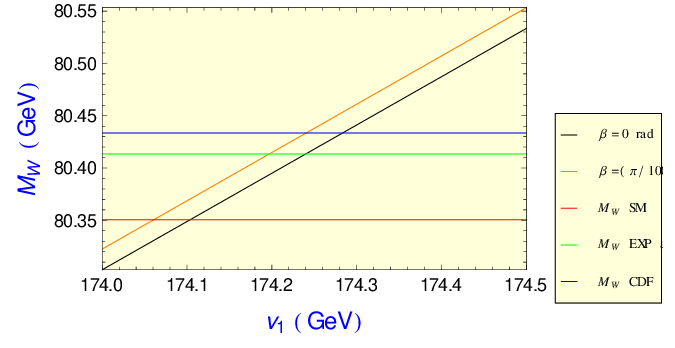}        
\caption{The masses of $W$ gauge boson in fuction of $v_{1}$ for $\theta =(\pi/4)$ rad, and some values of our $\beta$-parameter shown on the box. 
The value of line red is given by Eq.(\ref{smvalue}), the green by 
Eq.(\ref{average}) and the blue line for Eq.(\ref{cdfresult}).} 
\label{fgb3}
\end{figure}

\begin{figure}
\centering
\includegraphics[width=0.8\textwidth]{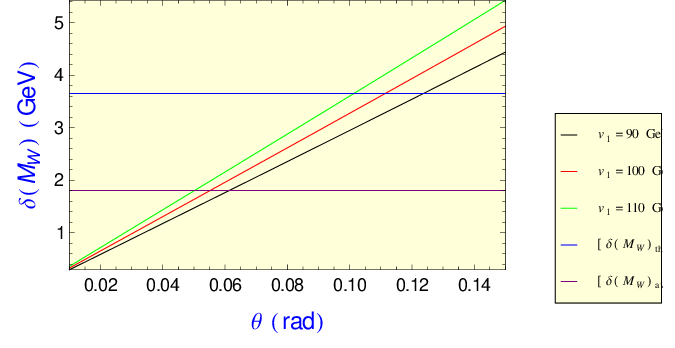}        
\caption{$\delta M_{W}$, defined in Eq.(\ref{wmassdifidentical}), as fuction 
of the new parameter $\theta$, the black 
line, for $v_{1} =90$ GeV; the red line $v_{1} =100$ GeV; the green line 
$v_{1} =110$ GeV to explain 
the values defined in our Eq.(\ref{deltamedidas}). This result is also 
hold in the case of the MSSM with $R$-Parity violation, see 
Eq.(\ref{wmassidentical}).} 
\label{dmwtheta}
\end{figure}

Our numerical results are shown at 
Figs.(\ref{fgb2},\ref{fgb3},\ref{dmwtheta}), where it is easy to 
convince yourself,  we can reproduce the experimental values of $W$ for several 
values of the parameters $\beta$ and $\theta$.

\begin{figure}
\centering
\includegraphics[width=0.8\textwidth]{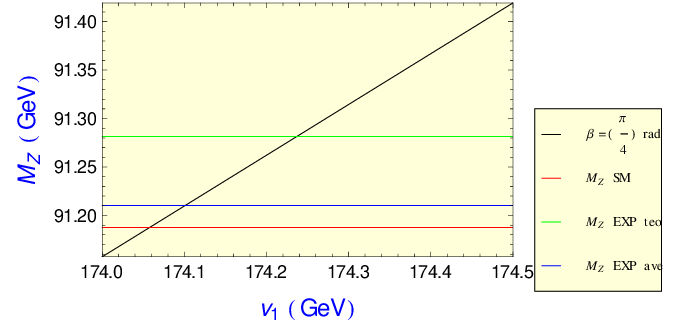}        
\caption{The masses of $Z$ gauge boson in fuction of $v_{1}$ and for 
$\beta = ( \pi /4)$ rad and $\zeta = tan^{-1} (5000/174.1)$, 
the red by Eq.(\ref{average}) and the green line 
for Eq.(\ref{averagethetadef}) while blue line for Eq.(\ref{cdfresultMZave}), see the first relation in our Eq.(\ref{zmassmssmrv}).} 
\label{mztreeidenticalb=45}
\end{figure}

\begin{figure}
\centering
\includegraphics[width=0.8\textwidth]{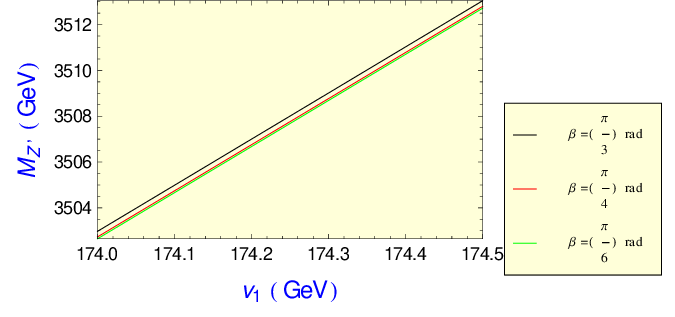}        
\caption{The masses of $Z^{\prime}$ gauge boson for $v_{1}$ and 
some  values of our $\beta$-parameter shown on the box, 
$\zeta = tan^{-1} (5000/174.1)$, see the second relation in our 
Eq.(\ref{zmassmssmrv}) and we see all the values are in agreement with 
experimental bound given by our Eq.(\ref{czzp}).} 
\label{mzptreeidenticalb=45}
\end{figure}

\begin{figure}
\centering
\includegraphics[width=0.8\textwidth]{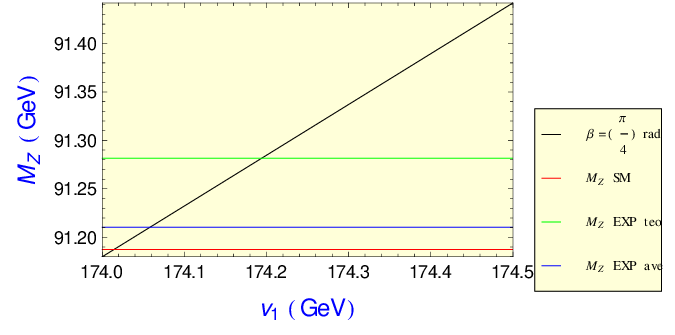}        
\caption{The masses of $Z$ gauge boson in fuction of $v_{1}$ for 
some  values of our $\beta$-parameter shown on the box, 
$\theta = ( \pi /100)$ rad, $\zeta = tan^{-1} (5000/174.1)$, 
the red by Eq.(\ref{average}) and the green line 
for Eq.(\ref{averagethetadef}) while blue line for Eq.(\ref{cdfresultMZave}), see the first relation in our Eq.(\ref{zmassmssmrv}).} 
\label{mztreeidenticalc=45}
\end{figure}

\begin{figure}
\centering
\includegraphics[width=0.8\textwidth]{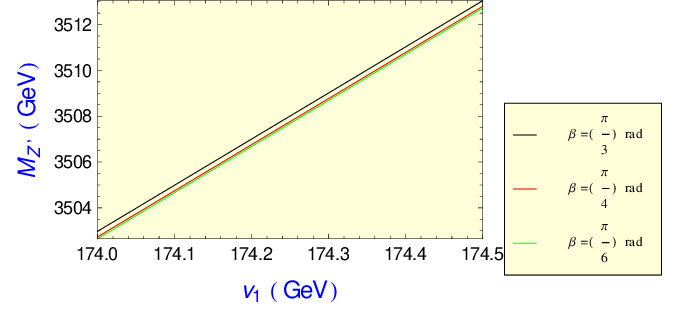}        
\caption{The masses of $Z^{\prime}$ gauge boson for $v_{1}=174.15$ GeV 
as fuction of $u_{2}$ for 
some values of $\xi$ parameter and some values of our $\beta$-parameter 
shown on the box, see the second relation in our 
Eq.(\ref{zmassmssmrv}) and we see all the values are in agreement with 
experimental bound given by our Eq.(\ref{czzp}).} 
\label{mzptreeidenticalc=45}
\end{figure}

The mass matrix for the neutal gauge boson is given by
\begin{eqnarray}
(M^{2})^{\rm{identical}}&=& \left( 
\begin{array}{ccc} 
A &- 
\frac{gg_{Y^{\prime}}}{4} 
\left[ \left( v_{1} \right)^{2}+ \left( v_{2} \right)^{2} \right] &- 
\frac{gg_{BL}}{4} \left( v^{L}_{3} \right)^{2}  \\
- 
\frac{gg_{Y^{\prime}}}{4} 
\left[ \left( v_{1} \right)^{2}+ \left( v_{2} \right)^{2} \right] & 
B &- g_{Y^{\prime}}g_{BL}\left( u^{2}_{1}+u^{2}_{2} \right) \\
- \frac{gg_{BL}}{4} \left( v^{L}_{3} \right)^{2} &-
g_{Y^{\prime}}g_{BL}\left( u^{2}_{1}+u^{2}_{2} \right) & C  
\end{array}
\right), \nonumber \\ 
\end{eqnarray}
where we have defined
\begin{eqnarray}
A&=&\frac{g^{2}}{4} \left[ \left( v_{1} \right)^{2}+ \left( v_{2} \right)^{2}+ \left( v^{L}_{3} \right)^{2} \right], \nonumber \\
B&=&g^{2}_{Y^{\prime}} 
\left( \frac{\left( v_{1} \right)^{2}+ \left( v_{2} \right)^{2}}{4}+ 
u^{2}_{1}+ u^{2}_{2} \right), \nonumber \\
C&=&g^{2}_{BL} 
\left( \frac{\left( v^{L}_{3} \right)^{2}}{4}+ 
u^{2}_{1}+ u^{2}_{2} \right) .
\end{eqnarray}
We can show the following analytical results
\begin{eqnarray}
{\mbox det}(M^{2})^{\rm{identical}}&=&0, \nonumber \\
{\mbox Tr}(M^{2})^{\rm{identical}}&=& A+B+C. \nonumber \\
\end{eqnarray}
Therefore we get the foton, and it is massless, and also two massive gauge 
boson $Z$ and $Z^{\prime}$ and their masses are
\begin{eqnarray}
M^{2}_{Z}&=&\frac{1}{8}\left[
\left(U+N \right) - 
\sqrt{ \left(U+N \right)^{2}- \left(V+O \right)} \right] , \nonumber \\
M^{2}_{Z^{\prime}}&=&\frac{1}{8}\left[
\left(U+N \right) + 
\sqrt{ \left(U+N \right)^{2}- \left(V+O \right)} \right] , \nonumber \\
\label{zmassmssmrv}
\end{eqnarray}
where we have defined and we also use Eq.(\ref{betadef})
\begin{eqnarray}
U&=&4\left(g_{Y^{\prime}}^{2}+g_{BL}^{2}\right)\left(u_{1}^{2}+u_{2}^{2}\right)
+\left(g^{2}+g_{Y^{\prime}}^{2}\right)\left(v_{1}^{2}+v_{2}^{2}\right) 
\nonumber \\
&=&4v_{1}^{2}\left(g_{Y^{\prime}}^{2}+g_{BL}^{2}\right)
\left(\tan^{2}\xi + tan^{2}\zeta \right) +v_{1}^{2}\left(g^{2}+g_{Y^{\prime}}^{2}\right)
\left(1+ \tan^{2}\beta \right),  \nonumber \\ 
N&=&\left( g^{2}+g^{2}_{BL}  \right)\left( v^{L}_{3} \right)^{2} =v_{1}^{2}\left( g^{2}+g^{2}_{BL}  \right) \tan^{2}\theta , \nonumber \\
V&=&16\left[g^{2}\left(g_{Y^{\prime}}^{2}+g_{BL}^{2}\right)+
g_{Y^{\prime}}^{2}g_{BL}^{2}\right]
\left(u_{1}^{2}+u_{2}^{2}\right)\left(v_{1}^{2}+v_{2}^{2}\right) \nonumber \\
&=&16v_{1}^{2}\left[g^{2}\left(g_{Y^{\prime}}^{2}+g_{BL}^{2}\right)+
g_{Y^{\prime}}^{2}g_{BL}^{2}\right] \left(1+ \tan^{2}\beta \right) 
\left( \tan^{2}\xi + tan^{2}\zeta \right), \nonumber \\
O&=&4\left[g^{2}\left(g_{Y^{\prime}}^{2}+g_{BL}^{2}\right)+
g_{Y^{\prime}}^{2}g_{BL}^{2}\right] \left(
v^{2}_{1}+v^{2}_{2}+4u^{2}_{1}+4u^{2}_{2} \right) 
\left( v^{L}_{3} \right)^{2} \nonumber \\
&=&4v^{2}_{1}\left[g^{2}\left(g_{Y^{\prime}}^{2}+g_{BL}^{2}\right)+
g_{Y^{\prime}}^{2}g_{BL}^{2}\right] 
\left( 1+ \tan^{2}\beta \right) 
\left( \tan^{2}\xi + tan^{2}\zeta \right), \nonumber \\
\label{massaapprox2}
\end{eqnarray}
where we have defined the new parameters $\xi$ and $\zeta$ in the following 
way:
\begin{eqnarray}
\tan \xi = \frac{u_{1}}{v_{1}}, \,\
\tan \zeta = \frac{u_{2}}{v_{1}}.
\end{eqnarray}
The trilinear gauge boson couplings are modified by the following mixing 
factor \cite{Pankov:2019yzr,Osland:2020onj} 
\begin{eqnarray}
\xi_{Z-Z^{\prime}}&=& \left( \frac{M_{Z}}{M_{Z^{\prime}}}
\right)^{2}, \nonumber \\
M_{Z^{\prime}}&>&3 \,\ {\mbox TeV},
\label{czzp}
\end{eqnarray}
this parameter, $\xi_{Z-Z^{\prime}}$, must be smaller than $10^{-3}$. On 
the other hand, we still have the following constraint \cite{Sanchez-Vega:2014rka}
\begin{equation}
\frac{M_{Z^{\prime}}}{g_{BL}}> 6 \,\ {\mbox TeV}.
\label{mzpdgbl}
\end{equation}
Our numerical result for this boson is presented in our 
Tab.(\ref{mzmzpczzp})  and 
Figs.(\ref{mztreeidenticalb=45},\ref{mzptreeidenticalb=45},\ref{mztreeidenticalc=45},\ref{mzptreeidenticalc=45}).

\begin{table}[h]
\begin{center}
\begin{tabular}{|c|c|c|c|c|}
\hline
$u_{2}$ (TeV) & $M_{Z}$ (GeV) & $M_{Z^{\prime}}$ (TeV) & 
$M_{Z^{\prime}}/g_{BL}$ (TeV) & $\xi_{Z-Z^{\prime}}$ \\
\hline
3.0 & 91.18 & 2.1 & 4.1 & 1.88 $\times 10^{-3}$  \\
\hline
3.5 & 91.18 & 2.5 & 4.8 & 1.38 $\times 10^{-3}$  \\
\hline
4.0 & 91.18 & 2.8 & 5.5 & 1.06 $\times 10^{-3}$  \\
\hline
4.5 & 91.18 & 3.2 & 6.2 & 8.4 $\times 10^{-4}$ \\
\hline
5.0 & 91.18 & 3.5 & 6.9 & 6.8 $\times 10^{-4}$  \\
\hline
5.5 & 91.18 & 3.9 & 7.6 & 5.6 $\times 10^{-4}$ \\
\hline
6.0 & 91.18 & 4.2 & 8.3 & 4.7 $\times 10^{-4}$  \\
\hline
6.5 & 91.18 & 4.6 & 9.0 & 4.0 $\times 10^{-4}$ \\
\hline
7.0 & 91.18 & 4.9 & 9.7 & 3.5 $\times 10^{-4}$ \\
\hline
\end{tabular}
\end{center}
\caption{The values for $M_{Z},M_{Z^{\prime}},(M_{Z^{\prime}}/g_{BL})$ and 
$\xi_{Z-Z^{\prime}}$ as function of $u_{2}$, for the case of 
$\beta = (\pi /4)$ rad, $\theta = (\pi /100)$ rad, $\xi =(\pi /3)$ rad and 
we satisfy the bonds of Eqs.(\ref{czzp},\ref{mzpdgbl}) when 
$u_{2}>4.5$ TeV.}
\label{mzmzpczzp}
\end{table}

The superpotential of this model have the following form \cite{Montero:2016qpx}
\begin{equation}
W= \mu_{H} \left( \hat{H}_{1}\hat{H}_{2} \right) 
+ \mu_{\phi}\hat{\phi}_{1}\hat{\phi}_{2}+ 
f^{l}_{ij} \left( \hat{H}_{2}\hat{L}_{i} \right) \hat{E}_{j}+
f^{\nu}_{ij} \left(\hat{H}_{1}\hat{L}_{i} \right) \hat{N}_{j}+
f^{M}_{ij}\hat{\phi}_{2}\hat{N}_{i}\hat{N}_{j},
\end{equation}
which generates mass to the model's charged leptons. We subsequently intend to analyze the masses of these leptons when $\tilde{L}_{3}$ acquires VEV.

\section{The Model with three non-identical Neutrinos}
\label{sec:noidentical}

The particle content of this model is as follows: we have the particles described in Eq.(\ref{leptonsm1}). In addition to these particles, 
replacing the three identical neutrinos defined by 
Eq.(\ref{righthandedneutrinosid}), we introducethree non-identical 
right-handed  neutrinos \cite{Montero:2016qpxno}
\begin{eqnarray}
&&
\hat{N}_{1R} \sim (1,{\bf 1},5,-5), \,\
\hat{N}_{\beta R} \sim (1,{\bf 1},-4,4) , \nonumber \\
\end{eqnarray}
where $\beta =2,3$. It is necessary to enlarge the scalar doublet sector adding four new scalars 
\begin{eqnarray}
\hat{\Phi}_{1} &=& \left(
\begin{array}{c}
\hat{\phi}^{0}_{1} \\
\hat{\phi}^{-}_{1}
\end{array}
\right) \sim (1,{\bf 2},5,-6), \,\
\hat{\Phi}^{\prime}_{1} = \left(
\begin{array}{c}
\hat{\phi}^{\prime +}_{1} \\
\hat{\phi}^{\prime 0}_{1}
\end{array}
\right) \sim (1,{\bf 2},-5,6), \nonumber \\
\hat{\Phi}_{2} &=& \left(
\begin{array}{c}
\hat{\phi}^{0}_{2} \\
\hat{\phi}^{-}_{2}
\end{array}
\right) \sim (1,{\bf 2},-4,3), \,\
\hat{\Phi}^{\prime}_{2} = \left(
\begin{array}{c}
\hat{\phi}^{\prime +}_{2} \\
\hat{\phi}^{\prime 0}_{2}
\end{array}
\right) \sim (1,{\bf 2},4,-3).
\label{extra3t}
\end{eqnarray}
In order to obtain an arbitrary mass matrix for the neutrinos we have to introduce the following additional singlet
\begin{eqnarray}
\hat{\varphi}_{1} \sim (1,{\bf 1},8,-8),  \,\
\hat{\varphi}_{2} \sim (1,{\bf 1},-10,10).
\label{singletsextra}
\end{eqnarray}

Using the VEV defined in our Eq.(\ref{vevyprcomum}) together with
\begin{eqnarray}
\varphi_{1}&=&\frac{1}{\sqrt{2}} \left(w_{1}+\textrm{Re}{\varphi}_{1}+ 
\imath \textrm{Im}{\varphi}_{1} \right),  \,\
\varphi_{2}= \frac{1}{\sqrt{2}} \left(w_{2}+\textrm{Re}{\varphi}_{2}+
\imath \textrm{Im}{\varphi}_{2}\right), \nonumber \\
\Phi_{1} &=& \left(
\begin{array}{c}
\frac{1}{\sqrt{2}} \left (u_{1}+ \textrm{Re}{\phi}_{1}^{0}+ 
\imath \textrm{Im} {\phi}_{1}^{0} \right) \\
\phi^{-}_{1}
\end{array}
\right), \,\
\Phi^{\prime}_{1} = \left(
\begin{array}{c}
\phi^{\prime +}_{1} \\
\frac{1}{\sqrt{2}} \left (u^{\prime}_{1}+ 
\textrm{Re}{\phi}^{\prime 0}_{1}+ 
\imath \textrm{Im} {\phi}^{\prime 0}_{1} \right)
\end{array}
\right), \nonumber \\
\Phi_{2} &=& \left(
\begin{array}{c}
\frac{1}{\sqrt{2}} \left (u_{2}+ \textrm{Re}{\phi}_{2}^{0}+ 
\imath \textrm{Im} {\phi}_{2}^{0} \right) \\
\phi^{-}_{2}
\end{array}
\right), \,\
\Phi^{\prime}_{2} = \left(
\begin{array}{c}
\phi^{\prime +}_{2} \\
\frac{1}{\sqrt{2}} \left (u^{\prime}_{2}+ 
\textrm{Re}{\phi}^{\prime 0}_{2}+ 
\imath \textrm{Im} {\phi}^{\prime 0}_{2} \right)
\end{array}
\right). \nonumber \\
\end{eqnarray}

The mass of charged gauge boson is 
\begin{eqnarray}
M^{2}_{W}&=& \frac{g^{2}}{4} \left[ \left( v_{1} \right)^{2}+ \left( v_{2} \right)^{2}+ \left( v^{L}_{3} \right)^{2}+ 
\left( u_{1} \right)^{2}+ \left( u_{2} \right)^{2}+ \left( u^{\prime}_{1} \right)^{2}+ \left( u^{\prime}_{2} \right)^{2} \right],
\label{wmassnoidentical}
\end{eqnarray}
now it is hold the following constraint
\begin{equation}
\sum_{i=1}^{2}
\left[
\left(v_{i} \right)^{2}+ \left( u_{i} \right)^{2}+ \left( u_{i}^{\prime} \right)^{2} + \left( v^{L}_{3} \right)^{2} 
\right] = (246)^{2} 
\,\ ({\mbox GeV})^{2}.
\end{equation}
To simplify our numerical analyses we will 
suppose 
\begin{eqnarray}
u_{1}&=&u_{2}=u, \nonumber \\
u^{\prime}_{1}&=&u^{\prime}_{2}=u^{\prime}, \nonumber \\
w_{1}&=&w_{2}=w,
\label{additionalsupos}
\end{eqnarray} 
in similar ways as done in \cite{Sanchez-Vega:2014rka}.

Using Eq.(\ref{additionalsupos}) and in similar way as done in MSSM \cite{Rodriguez:2019mwf,kazakov1}, we can 
rewrite this mass expression as
\begin{eqnarray}
M^{2}_{W}=\frac{g^{2}}{4}v^{2}_{1} \left[ 
\tan^{2} \beta +2 \tan^{2} \alpha +2 \tan^{2} \gamma + \tan^{2} \theta \right] \,\ , 
\label{wmassbetaparameter}
\end{eqnarray}
where the parameters $\beta$ and $\theta$ are defined in our Eq.(\ref{betadef}) and we have defined the following new free parameters:
\begin{eqnarray}
\tan \alpha = \frac{u}{v_{1}}, \,\ 
\tan \gamma = \frac{u^{\prime}}{v_{1}}.
\end{eqnarray}
We can also rewrite from Eq.(\ref{wmassbetaparameter}) the following relation
\begin{eqnarray}
\delta M_{W} &=& \frac{gv_{1}}{2}  
\sqrt{\tan^{2} \theta +2 \tan^{2} \alpha +2 \tan^{2} \gamma} .
\label{wmassdifnonidentical}
\end{eqnarray}
Ou results, again, is shown in our Fig.(\ref{dmwtheta}) when we take 
$\alpha = \gamma =0$ rad.

\begin{figure}[ht]
\begin{center}
\vglue -0.009cm
\mbox{\epsfig{file=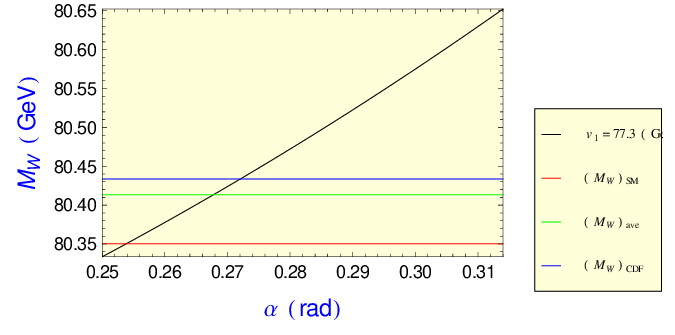,width=0.7\textwidth,angle=0}}       
\end{center}
\caption{The Masses of $W$- Gauge Bosons, in GeV in function of 
$\alpha$-parameter, see Eq.(\ref{wmassbetaparameter}), for 
$v_{1}=77.3$ GeV and $\beta = \gamma = \delta = ( \pi/3)$ rad and 
$\theta = ( \pi/100)$ rad.}
\label{fig2}
\end{figure}

In Fig.(\ref{fig2}) we shown the 
behaviour of $M_{W}$ as function of $\alpha$-parameter and as conclusion 
we can explain in easy way this experimental values defined in ours 
Eqs.(\ref{smvalue},\ref{average},\ref{cdfresult}). 

The mass matrix for the neutal gauge boson, in the base $\left( W^{3}_{m},(b_{Y^{\prime}})_{m},(b_{BL})_{m} \right)$, 
is given by \cite{Sanchez-Vega:2014rka}
\begin{eqnarray}
M^{2}_{neutral}= \left(
\begin{array}{ccc}
g^{2}(K+P+2N)) & 
-gg_{Y^{\prime}}(K+N) & 
-gg_{BL}(P+N) \\
-gg_{Y^{\prime}}(K+N) & 
g_{Y^{\prime}}^{2}K & 
g_{Y^{\prime}}g_{BL}N \\
-gg_{BL}(P+N) & 
g_{Y^{\prime}}g_{BL}N & g_{BL}^{2}P
\end{array}
\right) \nonumber \\
\label{massmatrixneutralgeralbruce} 
\end{eqnarray}
where
\begin{eqnarray}
K&=& \frac{1}{4} \sum_{a} V^{2}_{a}(Y^{\prime}_{a})^{2} 
= \frac{v^{2}_{2}}{4}\left[ 1+ \tan^{2}\beta +41 \left( \tan^{2}\alpha + \tan^{2}\gamma \right) +164 \tan^{2}\delta 
\right] , \nonumber \\
P&=& \frac{1}{4} \sum_{a}V^{2}_{a}[(B-L)_{a}]^{2} 
= \frac{v^{2}_{2}}{4}\left[ \tan^{2}\theta +45 \left( \tan^{2}\alpha + \tan^{2}\gamma \right) +164 \tan^{2}\delta 
\right] , \nonumber \\
N&=& \frac{1}{4} \sum_{a}V^{2}_{a}(Y^{\prime}_{a})(B-L)_{a} 
=- \frac{v^{2}_{2}}{4}\left[ 42 \left( \tan^{2}\alpha + \tan^{2}\gamma \right) +164 \tan^{2}\delta 
\right]; \nonumber \\
\end{eqnarray}
where $g,g_{Y^{\prime}}$ and $g_{BL}$ are the gauge coupling constants are defined in 
the Tab.(\ref{table2}). The numbers $(Y^{\prime}_{a})$ and $(B-L)_{a}$ being 
the quantum numbers defined in Tab.(\ref{lepm1}) and the label $a$ means 
all scalars of this model.

\begin{figure}[ht]
\begin{center}
\vglue -0.009cm
\mbox{\epsfig{file=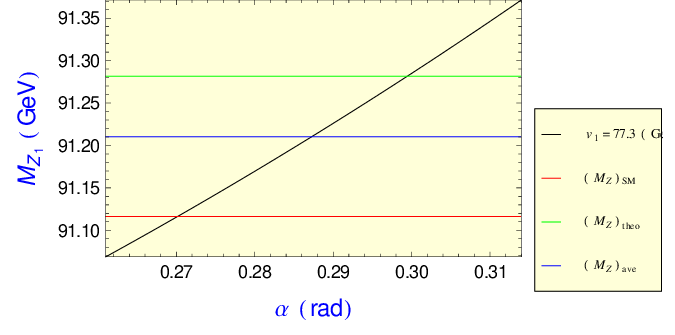,width=0.7\textwidth,angle=0}}       
\end{center}
\caption{The Masses of $Z_{1}$-Neutral Gauge Bosons, in GeV as function of 
$v_{1}$, for $\beta = ( \pi/3 )$ rad, $\theta = ( \pi/100 )$ rad, 
$\gamma = \delta = ( \pi/3 )$ rad, see Eq.(\ref{masseigestatesgeral}).}
\label{figmz1}
\end{figure}

\begin{figure}[ht]
\begin{center}
\vglue -0.009cm
\mbox{\epsfig{file=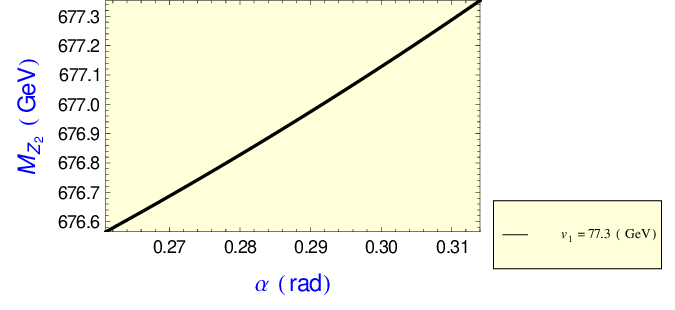,width=0.7\textwidth,angle=0}}       
\end{center}
\caption{The Masses of $Z_{2}$-Neutral Gauge Bosons, in GeV as function of 
$v_{1}$, for $\beta = ( \pi/3 )$ rad, $\theta = ( \pi/100 )$ rad, 
$\gamma = \delta = ( \pi/3 )$ rad, see Eq.(\ref{masseigestatesgeral}) 
and we see all the values are in agreement with 
experimental bound given by our Eq.(\ref{czzp}).}
\label{figmz2}
\end{figure}

\begin{figure}[ht]
\begin{center}
\vglue -0.009cm
\mbox{\epsfig{file=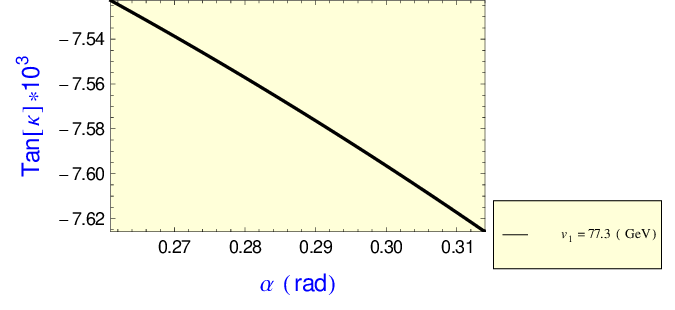,width=0.7\textwidth,angle=0}}       
\end{center}
\caption{The $\kappa$-parameter as function of $\alpha$-parameter in rad, 
for $\beta = ( \pi/3 )$ rad, $\theta = ( \pi/100 )$ rad, 
$\gamma = \delta = ( \pi/3 )$ rad, see Eq.(\ref{defkappaprime}).}
\label{figtH}
\end{figure}

It is easy to show
\begin{eqnarray}
{\mbox Det}\left( M^{2}_{{\mbox neutral}}\right)&=&0, 
\nonumber \\
{\mbox Tr}\left(M^{2}_{{\mbox neutral}}\right)&=&
g^{2}(K+P+2N)+g_{Y^{\prime}}^{2}K+
g_{BL}^{2}P \equiv R,
\label{Rdef}
\end{eqnarray} 
therefore we have one non massive Gauge 
Boson and it is the photon $A_{m}$ plus 
two massive eigenstates, we will call them as $Z_{1m}$, $Z_{2m}$. 
By another hand the characteristic equation for this mass matrix is
\begin{equation}
x \left( -T+Rx-x^{2} \right) \equiv 0,
\end{equation}
where $R$ is defined in our Eq.(\ref{Rdef}) and we define $T$ in the 
following way
\begin{equation}
T=(KP-N^{2})
\left[ g^{2}(g_{Y^{\prime}}^{2}+g_{BL}^{2})+g_{Y^{\prime}}^{2}g_{BL}^{2} 
\right].
\end{equation}
The massives eigenvalues are given by \cite{Sanchez-Vega:2014rka}:
\begin{eqnarray}
M^{2}_{Z_{1}}&=&\frac{1}{2} \left[ R - \sqrt{ R^{2}-4T} \right], 
\nonumber \\
M^{2}_{Z_{2}}&=&\frac{1}{2} \left[ R + \sqrt{ R^{2}-4T} \right], 
\nonumber \\
\label{masseigestatesgeral}
\end{eqnarray}
and the mass eigenstates\footnote{To understand the eigenvalues see our App.(\ref{partialresultsnonidentical}).} \cite{Sanchez-Vega:2014rka}:
\begin{eqnarray}
A_{m}&=&\frac{1}{N_{\gamma}} \left[ 
\frac{1}{g}W^{3}_{m}+ \frac{1}{g_{Y^{\prime}}}(b_{Y^{\prime}})_{m}+ 
\frac{1}{g_{BL}}(b_{BL})_{m} \right], \nonumber \\
Z_{m}&=& \cos \theta_{W}W^{3}_{m}- \sin \theta_{W} \sin \vartheta 
(b_{Y^{\prime}})_{m}- \sin \theta_{W} \cos \vartheta(b_{BL})_{m}, 
\nonumber \\
Z^{\prime}_{m}&=&\cos \vartheta (b_{Y^{\prime}})_{m}- 
\sin \vartheta (b_{BL})_{m}, \nonumber \\
\end{eqnarray}
where the $\vartheta$-parameter is defined at 
Eq.(\ref{defalphaprime}). We can write the two physical massive 
gauge bosons $Z_{1}$ and $Z_{2}$ in terms of 
the usual no physical gauge bosons $Z$ and $Z^{\prime}$ in the following way \cite{Sanchez-Vega:2014rka}:
\begin{eqnarray}
(Z_{1})_{m}&=& \cos \kappa Z_{m}+ \sin \kappa Z^{\prime}_{m}, \nonumber \\
(Z_{2})_{m}&=&- \sin \kappa Z_{m}+ \cos \kappa Z^{\prime}_{m},
\end{eqnarray}
where we have defined
\begin{equation}
\tan \kappa \equiv 
\frac{\sqrt{g^{2}(g^{2}_{BL}+g^{2}_{Y^{\prime}})+g^{2}_{BL}g^{2}_{Y^{\prime}}}(g^{2}_{BL}P-g^{2}_{Y^{\prime}}N-M^{2}_{Z_{2}})}{g^{2}(g^{2}_{BL}+g^{2}_{Y^{\prime}})(P+N)+g^{2}_{Y^{\prime}}(g^{2}_{BL}(P+N)-M^{2}_{Z_{2}})}.
\label{defkappaprime}
\end{equation}
Our numerical results are shown in our Figs.(\ref{figmz1},\ref{figmz2},\ref{figtH}) where we can say
\begin{itemize}
\item $Z_{1m}$ $\simeq$ $Z_{m}$;
\item $Z_{2m}$ $\simeq$ $Z^{\prime}_{m}$;
\item $\tan \kappa$ $\simeq$ $10^{-3}$;
\end{itemize}
those resultas are in agreement with \cite{Sanchez-Vega:2014rka}.

The superpotential of this model have the following form \cite{Montero:2016qpxno}
\begin{eqnarray}
W&=& \mu_{H} \left( \hat{H}_{1}\hat{H}_{2} \right) + 
\mu_{\Phi_{1}} \left( \hat{\Phi}^{\prime}_{1}\hat{\Phi}_{1} \right) +
\mu_{\Phi_{2}} \left( \hat{\Phi}^{\prime}_{2}\hat{\Phi}_{2} \right) +
f^{l}_{ij} \left( \hat{H}_{2}\hat{L}_{i} \right) \hat{E}_{j} 
\nonumber \\ &+&
f^{\nu}_{i} \left(\hat{\Phi}^{\prime}_{1}\hat{L}_{i} \right) \hat{N}_{1}+
f^{\nu}_{i \beta} \left(\hat{\Phi}^{\prime}_{2}\hat{L}_{i} \right) 
\hat{N}_{\beta}+
f^{M} \hat{\varphi}_{2}\hat{N}_{1} \hat{N}_{1}+
f^{M}_{\alpha \beta} \hat{\varphi}_{2}\hat{N}_{\alpha}\hat{N}_{\beta},
\end{eqnarray}
which generates mass to the model's charged leptons. We subsequently intend to analyze the masses of these leptons when $\tilde{L}_{3}$ acquires VEV.

\section{Conclusions}
\label{sec:con}
We have studied the masses of gauge boson masses in two supersymmetric model, 
first with three identical right-handed neutrinos and then for the case with 
three noidentical right-handed neutrinos. We can explain the recent CDF 
measure of the $W$-boson mass in both models for several values of our 
$\beta$ and $\theta$ parameters, as shown in our Fig.(\ref{dmwtheta}), 
this results is hold for the MSSM with $R$-Parity violation, see 
Eq.(\ref{deltamedidas}), as well for 
the model with three-idendical neutrinos, see Eq.(\ref{wmassdifidentical}), 
and also for three non identical three right handed neutrinos, 
Eq.(\ref{wmassdifnonidentical}). Our results for neutral gauge bosons 
are given in Figs.(\ref{mztreeidenticalb=45},\ref{mzptreeidenticalb=45},\ref{mztreeidenticalc=45},\ref{mzptreeidenticalc=45}) for the case of three 
identical neutrinos and in Figs.(\ref{figmz1},\ref{figmz2}) for the three 
non-identical right and our results are in agreement with current 
experimental limits. Next we need to study the masses of the leptons 
as well the masses of the scalar sectors of thoses models.

\begin{center}
{\bf Acknowledgments} 
\end{center}
We would like thanks V. Pleitez, J. C. Montero and B. L. Sch\'anchez-Vega for 
useful discussions above the non-supersymmetric version of this model. 
We also want to thank IFT for the nice hospitality during my several visit 
to perform my studies about the severals Models with gauge symmetry 
$U(1)_{Y^\prime}\times U(1)_{B-L}$ and also for done 
this article.

\appendix

\section{Minimal Supersymmetric Standard Model with $R$-Parity violation}
\label{sec:mssmrpv}

In this model, as in MSSM with $R$-parity conservation 
\cite{Rodriguez:2019mwf,kazakov1}, we will introduce the following 
chiral superfields for the leptons and usual scalars fields 
\cite{Rodriguez:2022gfq}
\begin{eqnarray}
\hat{L}_{iL} &=& \left( 
\begin{array}{c}  
\hat{\nu}_{iL} \\
\hat{l}_{iL}
\end{array}
\right) \sim \left( 1,{\bf 2},-1 \right), \,\ 
\hat{E}_{iR}\sim \left( 1,{\bf 1},+2 \right), \,\ i=1,2,3, \nonumber \\
\hat{H}_{1} &=& \left( 
\begin{array}{c}  
\hat{H}^{+}_{1} \\
\hat{H}^{0}_{1}
\end{array}
\right) \sim \left( 1,{\bf \bar{2}},+1 \right), \,\ 
\hat{H}_{2} = \left( 
\begin{array}{c}  
\hat{H}^{0}_{2} \\
\hat{H}^{-}_{2}
\end{array}
\right) \sim \left( 1,{\bf 2},-1 \right).
\end{eqnarray}
Remember in each chiral superfield we have both spin one-half, fermions and 
Higgsinos, beyond the scalars, sfermions and Higgs fields. Therefore, the 
left-handed and right-handed fermions fields have different representation 
of the gauge group as in the SM. The neutral scalars obtain the 
VEV defined in our Eq.(\ref{vevyprcomum}). 

Concerning the gauge bosons and their superpartners, they are introduced in vector superfields. See Table~\ref{tablemssmrpv} the particle content together with the gauge coupling constant of each group.

\begin{table}[h]
\begin{center}
\begin{tabular}{|c|c|c|c|c|}
\hline
${\rm{Group}}$ & ${\rm Vector \,\ Superfield}$ & ${\rm{Gauge \,\ Bosons}}$ & ${\rm{Gaugino}}$ & ${\rm Gauge \,\ constant}$ \\
\hline
$SU(3)_{C}$ & $\hat{G}^{a}$ & $g^{a}_{m}$ & $\tilde{g}^{a}$ & $g_{s}$  \\
\hline
$SU(2)_{L}$ & $\hat{W}^{i}$ & $W^{i}_{m}$ & $\tilde{W}^{i}$ & $g$ \\
\hline
$U(1)_{Y}$ & $\hat{b}$ & $b_{m}$ & $\tilde{b}$ & $g_{Y}$ \\
\hline
\end{tabular}
\end{center}
\caption{Information on fields contents of each vector superfield of this model, see our Tab.(\ref{tablegaugebosonssm}).}
\label{tablemssmrpv}
\end{table}

The coupling constant $g$ is given by
\begin{equation}
g= \sqrt{\frac{8G_{F}M^{2}_{W}}{\sqrt{2}}}=0.652673,
\end{equation}
where the Fermi constant is
\begin{equation}
G_{F}=1.1663787 \times 10^{-5} \,\ {\mbox GeV}^{-2},
\end{equation}
the vaccum expectation value can be estimate using the following relation
\begin{equation}
v= \frac{1}{\sqrt{\sqrt{2}G_{F}}}\simeq 246.22 \,\ {\mbox GeV}.
\label{extvevsmfermi}
\end{equation}

We get the following expression to the charged gauge bosons masses 
\begin{eqnarray}
M^{2}_{W}&=& \frac{g^{2}}{4} \left[ v_{1}^{2}+v_{2}^{2}+ \left( v^{L}_{3} \right)^{2} \right] =
\frac{g^{2}v_{1}^{2}}{4} \left[ 1+ \tan^{2} \beta + \tan^{2} \theta \right].
\label{wmassrpv}
\end{eqnarray}
see Eqs.(\ref{wmassidentical},\ref{betadef}). 

The mass matrix for the neutal gauge boson is
\begin{eqnarray}
(M^{2})^{\mbox{RPV}}&=& \frac{g^{2}v_{1}^{2}}{4} 
\left[ 1+ \tan^{2} \beta + \tan^{2} \theta \right] 
\left( 
\begin{array}{cc}  
1 &- \tan \theta_{W}  \\
- \tan \theta_{W} & \tan^{2} \theta_{W}
\end{array}
\right),
\label{neutralgaugemssmrpv} 
\end{eqnarray}
where $\theta_{W}$ is the Weinberg angle and it is defined as
\begin{eqnarray}
\tan \theta_{W} &\equiv& \frac{g_{Y}}{g}, \nonumber \\
e&=&g \sin \theta_{W}=g_{Y}\cos \theta_{W}, \nonumber \\
g_{Y}&=&g \left( \frac{M_{Z}}{M_{W}} \right) 
\sqrt{1- \left( \frac{M_{W}}{M_{Z}} \right)^{2}} \simeq 0.350221.
\label{gySM}
\end{eqnarray} 

\begin{figure}
\centering
\includegraphics[width=0.8\textwidth]{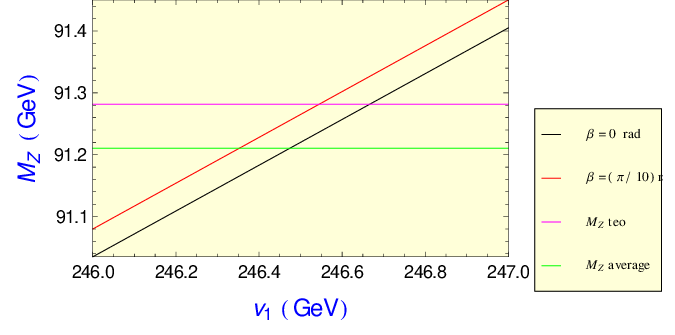}        
\caption{The masses of $Z$ gauge boson in fuction of 
$v_{1}$ for $\theta =0$ rad and for some values of our 
$\beta$-parameter shown on the box, the red line, 
the green by Eq.(\ref{average}) and the purple line and the 
green line are defined for Eq.(\ref{cdfresultMZave}).} 
\label{mzvsv1}
\end{figure}

We can show, from Eq.(\ref{neutralgaugemssmrpv}), the following 
analytical results
\begin{eqnarray}
{\mbox det}(M^{2})^{\mbox{RPV}}&=&0, \nonumber \\
{\mbox Tr}(M^{2})^{\mbox{RPV}}&=& \frac{g^{2}v_{1}^{2}}{4 \cos^{2} \theta_{W}} 
\left[ 1+ \tan^{2} \beta + \tan^{2} \theta \right]= 
\frac{M^{2}_{W}}{\cos^{2} \theta_{W}}. \nonumber \\
\end{eqnarray}
Therefore we get the foton, and it is massless, and also a massive gauge 
boson $Z$ and its masses is
\begin{eqnarray}
M^{2}_{Z}&=& \frac{M^{2}_{W}}{\cos^{2} \theta_{W}}, \nonumber \\
\label{zmassmssmrvrpv}
\end{eqnarray}
and our numerical result for this boson is presented in our 
Tab.(\ref{masswandzmssmrpv}) and in our Fig.(\ref{mzvsv1}).

\begin{table}[h]
\begin{center}
\begin{tabular}{|c|c|c|}
\hline
$v_{1} \mbox{(GeV)}$ &  $M_{W} \mbox{(GeV)}$ & $M_{Z} \mbox{(GeV)}$ \\ \hline
246.21 & 80.3474 & 91.1127 \\ \hline
246.211 & 80.3477 & 91.1131 \\ \hline
256.212 & 80.3480 & 91.1135 \\ \hline
256.213 & 80.3483 & 91.1138 \\ \hline
256.214 & 80.3487 & 91.1142 \\ \hline
256.215 & 80.3490 & 91.1146 \\ \hline
256.216 & 80.3493 & 91.1149 \\ \hline
256.217 & 80.3496 & 91.1153 \\ \hline
256.218 & 80.3500 & 91.1157 \\ \hline
256.219 & 80.3503 & 91.1160 \\ \hline
256.220 & 80.3506 & 91.1164 \\ \hline
\end{tabular}
\end{center}
\caption{The masses of $W$- and $Z$-bosons, see our 
Eqs.(\ref{wmassrpv},\ref{zmassmssmrvrpv}), when $\beta = \theta =0$rad when 
we recover the SM formulas given in our 
Eqs.(\ref{wmasssm},\ref{zmasssm}). See we get the same $v$ as defined in 
Eq.(\ref{extvevsmfermi}).}
\label{masswandzmssmrpv}
\end{table}

The neutral massive gauge boson $Z$
\begin{eqnarray}
M_{Z}= \frac{M_{W}}{\cos \theta_{W}},
\label{zmasssm}
\end{eqnarray}
the average value of experimental 
measurements provides us with the following data \cite{average}
\begin{eqnarray}
\left( M_{Z} \right)_{\rm EXP}&=& \left(
91.1875 \pm 0.0021 \right) \,\ {\mbox GeV}, \nonumber \\
\left( \sin \theta^{lep}_{eff}(Q^{had}_{FB}) \right)&=& 
0.2324 \pm 0.0012, \,\
\left( \sin \theta^{lep}_{eff}(HC) \right)= 0.23143 \pm 0.00025. \nonumber \\
\label{averagemz}
\end{eqnarray}
where $\theta_{W}$ is the Weinberg angle and it is defined  in our 
Eq.(\ref{gySM}) and those values are in agreement with
\begin{eqnarray}
\left( \cos^{2} \theta_{W} \right)_{\rm EXP}= 
\frac{\left( M_{W} \right)^{2}_{\rm EXP}}{\left( M_{Z} \right)^{2}_{\rm EXP}} 
=0.778, \,\ 
\left( \sin^{2} \theta_{W} \right)= 0.222.
\label{averagethetadef}
\end{eqnarray}

The result presented in our 
Eq.(\ref{cdfresult}) together Eq.(\ref{zmasssm}) will imply
\begin{eqnarray}
\left( M_{Z} \right)_{\rm CDF}&=& \frac{\left( M_{W} \right)_{\rm CDF}}{ \cos \theta_{W}}= 91.21  \,\ {\mbox GeV}.
\label{cdfresultMZ}
\end{eqnarray}

We can define
\begin{eqnarray}
\left( \delta M^{2}_{W} \right)^{theo}&=& 
\left( M^{2}_{W} \right)_{\rm CDF}- \left( M^{2}_{W} \right)_{\rm SM}
=13.3451 \,\ {\mbox GeV}^{2}, \nonumber \\
\left( \delta M^{2}_{W} \right)^{ave}&=& 
\left( M^{2}_{W} \right)_{\rm CDF}- \left( M^{2}_{W} \right)_{\rm EXP}
=3.2491 \,\ {\mbox GeV}^{2}, \nonumber \\
\left( \delta M_{Z} \right)^{theo}&=& 
\frac{\left( \delta M_{W} \right)^{theo}}{\left( \cos \theta_{W} \right)_{\rm EXP}}=4.14 \,\ {\mbox GeV}, 
\,\
\left( \delta M_{Z} \right)^{ave}= 
\frac{\left( \delta M_{W} \right)^{ave}}{\left( \cos \theta_{W} \right)_{\rm EXP}}=2.04 \,\ {\mbox GeV}. \nonumber \\
\label{deltamedidas}
\end{eqnarray}
those two last relation imply
\begin{eqnarray}
\left( M^{teo}_{Z} \right)_{\rm CDF}&=& 91.28  \,\ {\mbox GeV}, \,\
\left( M^{ave}_{Z} \right)_{\rm CDF}= 91.21  \,\ {\mbox GeV},
\label{cdfresultMZave}
\end{eqnarray}
its value is the same as given by our Eq.(\ref{cdfresultMZ}).

\begin{figure}
\centering
\includegraphics[width=0.8\textwidth]{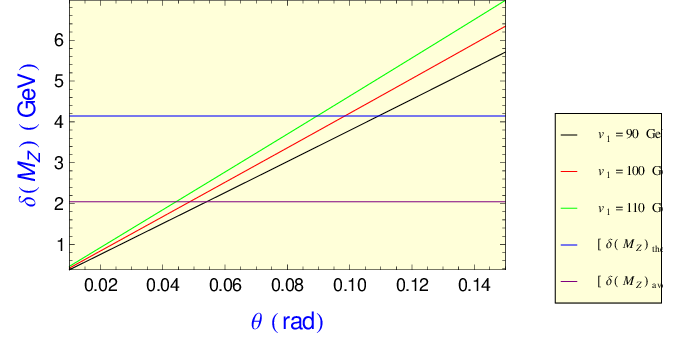}        
\caption{$\delta M_{Z}$ in fuction of the new parameter $\theta$, the black 
line, for $v_{1} =90$ GeV; $v_{1} =100$ GeV; $v_{1} =110$ GeV to explain 
the values defined in our Eq.(\ref{deltamedidas}).} 
\label{dmztheta}
\end{figure}

We can write from our Eqs.(\ref{wmassrpv},\ref{zmassmssmrv}) we can 
show
\begin{eqnarray}
\delta M_{W}&=& \frac{gv_{1}}{2}\tan \theta , \,\
\delta M_{Z}= \frac{\delta M_{W}}{\cos \theta_{W}}, \nonumber \\
\label{deltamedidas}
\end{eqnarray}
our results for those parameters are shown in our 
Figs.(\ref{dmwtheta},\ref{dmztheta}), where we shown we can accomodate the 
measurement of $W$-boson mass for several values of $v_{1}$ when we fix 
the new parameter $\theta$.

The superpotential that violate $R$-parity and define MSSMRV is given by 
\begin{eqnarray}
W_{MSSMRV}&=& W_{2}+ W_{3RC},   \\
W_{2}&=&\mu\; \left( \hat{H}_{1}\hat{H}_{2} \right)+ \sum_{i=1}^{3}\mu_{0i} \left( \hat{L}_{i}\hat{H}_{2} \right),\nonumber \\ 
W_{3}&=& \sum_{i,j,k=1}^{3}\left[\, 
f^{l}_{ij}\left( \hat{H}_{1}\hat{L}_{i}\right) \hat{E}_{jR}+
f^{d}_{ij}\left( \hat{H}_{1}\hat{Q}_{i}\right) \hat{D}_{jR}+
f^{u}_{ij}\left( \hat{H}_{2}\hat{Q}_{i}\right) \hat{U}_{jR}+
\lambda_{ijk} \left( \hat{L}_{i}\hat{L}_{j}\right) \hat{E}_{kR}
\right. \nonumber \\ &+& \left.
\lambda^{\prime}_{ijk}\left( \hat{L}_{i}\hat{Q}_{j}\right) \hat{D}_{kR} \right].
\label{suppotMSSM}
\end{eqnarray}

\section{To understand the mixing in neutral gauge bosons in the model with three noidentical neutrinos}
\label{partialresultsnonidentical}

In this model we recover the Standard Model when
\begin{equation}
\beta = \theta = \alpha = \gamma = \delta =0, \,\ \mbox{rad},
\end{equation}
in this case our Eq.(\ref{massmatrixneutralgeralbruce}) is given by
\begin{eqnarray}
M^{2}&=& \frac{g^{2}v^{2}}{4} \left( 
\begin{array}{ccc} 
1 &- \left( \frac{g_{Y^{\prime}}}{g} \right) & 0  \\
- \left( \frac{g_{Y^{\prime}}}{g} \right) & 
\left( \frac{g_{Y^{\prime}}}{g} \right)^{2} & 0 \\
0 & 0 & 0  
\end{array}
\right), \nonumber \\
\label{massmatrixneutralsm} 
\end{eqnarray}
this mass matrix in the superior block $2 \times 2$ is very similar mass 
matrix of the Standard Model where we exchange 
$g \rightarrow g_{Y^{\prime}}$. The eigenvalues for this mass matrix 
are\footnote{See our Eq(\ref{masseigestatesgeral}).}
\begin{eqnarray}
M^{2}_{Z_{1}}&=&0, \nonumber \\
M^{2}_{Z_{2}}&=&\frac{g^{2}v^{2}}{4}\left[ 
1+ \left( \frac{g_{Y^{\prime}}}{g} \right)^{2}
\right], \nonumber \\
\label{masszezpsm}
\end{eqnarray}
and we can conclude
\begin{eqnarray}
Z_{m}&=& \cos \theta_{W}W^{3}_{m}+ c_{1} \sin \theta_{W} b_{Y^{\prime}})_{m}, \nonumber \\
Z^{\prime}_{m}&=&c_{2} (b_{Y^{\prime}})_{m}+c_{3}(b_{BL})_{m}, 
\end{eqnarray}
where $c_{1},c_{2}$ and $c_{3}$ are the coefficients responsible in mix $b_{Y^{\prime}})_{m}$ with $(b_{BL})_{m}$.

In this model we recover the MSSM with $R$ Parity conservation when
\begin{equation}
\theta = \alpha = \gamma = \delta =0, \,\ \mbox{rad},
\end{equation}
in this case, we again get our Eq.(\ref{massmatrixneutralgeralbruce}) but in 
this case our Eq.(\ref{masszezpsm}) become
\begin{eqnarray}
M^{2}_{Z_{1}}&=&0, \nonumber \\
M^{2}_{Z_{2}}&=&\frac{g^{2}v^{2}}{4}\left[ 
1+ \left( \frac{g_{Y^{\prime}}}{g} \right)^{2}
\right] \left( 1+ \tan^{2} \beta \right). \nonumber \\
\label{masszezpmssm}
\end{eqnarray}

In this model we recover the Standard Model with $R$ Parity violation when 
Eq.(\ref{zmassmssmrvrpv}) is hold, 
in this case our Eq.(\ref{massmatrixneutralgeralbruce}) become
\begin{eqnarray}
M^{2}&=& \frac{g^{2}v^{2}}{4} \left( 
\begin{array}{ccc} 
\left( 1+ \tan^{2}\beta \right) &- \left( \frac{g_{Y^{\prime}}}{g} \right) \left( 1+ \tan^{2}\beta \right) & 0  \\
- \left( \frac{g_{Y^{\prime}}}{g} \right) \left( 1+ \tan^{2}\beta \right) & 
\left( \frac{g_{Y^{\prime}}}{g} \right)^{2}\left( 1+ \tan^{2}\beta \right) & 0 \\
0 & 0 & 0  
\end{array}
\right). \nonumber \\
\label{massmatrixneutralmssmrpc} 
\end{eqnarray}


\begin{thebibliography}{99}
\bibitem{sg}S. J. L. Rosner,
\emph{Resource letter: The Standard model and beyond},
\emph{ Am. J. Phys.} {\bf 71}, 302, (2003).
\bibitem{Kronfeld:2010bx} A. S. Kronfeld and C. Quigg,
\emph{Resource Letter: Quantum Chromodynamics},
\emph{ Am. J. Phys.}{\bf78}, 1081, (2010).

\bibitem{average}J. de Blas, M. Pierini, L. Reina and L. Silvestrini, 
\emph{Impact of the recent measurements of the top-quark and $W$-boson masses on electroweak precisions fits}, 
\emph{Phys. Rev. Lett.}{\bf 129}, 271801, (2022).
\bibitem{cdf}T. Aaltonen {\it et al} (CDF Collaboration), 
\emph{High-precision measurement of the $W$ boson mass with the CDF II 
detector}, 
\emph{ Science} {\bf 376}, 170, (2022).
\bibitem{Rodriguez:2022hsj}
M. C. Rodriguez,
{\sl Gauge bosons masses in the context of the Supersymmetric $SU(3)_{C}\otimes SU(3)_{L}\otimes U(1)_{N}$ Model}, [arXiv:2209.04653 [hep-ph]].
\bibitem{Blank:1997qa}T. Blank and W. Hollik,
{\it Precision observables in SU(2) x U(1) models with an additional Higgs triplet},
{\sl Nucl. Phys.}{\bf B514}, 113-134, (1998);
[arXiv:hep-ph/9703392 [hep-ph]].
\bibitem{Chen:2006pb}M. C. Chen, S. Dawson and T. Krupovnickas,
{\it Higgs triplets and limits from precision measurements},
{\sl Phys. Rev.}{\bf D74}, 035001, (2006);
[arXiv:hep-ph/0604102 [hep-ph]].
\bibitem{evans}J. L. Evans, T. T. Yanagida and N. Yokozaki,
{\it W boson mass anomaly and grand unification},
[arXiv:2205.03887] [hep-ph].
\bibitem{Bhaskar:2022vgk} A. Bhaskar, A. A. Madathil, T. Mandal and S. Mitra,
{\it Combined explanation of $W$-mass, muon $g-2$, $R_{K^{(*)}}$ and $R_{D^{(*)}}$ anomalies in a singlet-triplet scalar leptoquark model},
[arXiv:2204.09031 [hep-ph]].
\bibitem{Kribs:2007ac} G. D. Kribs, E. Poppitz, and N. Weiner, 
{\it Flavor in supersymmetry with an extended R-symmetry},
{\sl Phys. Rev.}{\bf D78}, 055010, (2008); arXiv:0712.2039 [hep-ph].
\bibitem{Diessner:2014ksa}P. Die\ss{}ner, J. Kalinowski, W. Kotlarski, and D. St\"ockinger, 
{\it Higgs boson mass and electroweak observables in the MRSSM},
{\sl JHEP} {\bf 12}, 124, (2014); arXiv:1410.4791 [hep-ph].
\bibitem{Diessner:2019ebm} P. Diessner and G. Weiglein, 
{\it Precise prediction for the W boson mass in the MRSSM}, 
{\sl JHEP}{\bf 07}, 011, (2019); arXiv:1904.03634 [hep-ph].
\bibitem{athron}P. Athron, M. Bach, D. H. J. Jacon, W. Kotlarski, D. 
St\"ockinger and A. Voigt, {\it Precise calculation of the $W$ boson pole 
mass beyond the Standard Model with FlexibleSUSY}, [arXiv:2204.05285] [hep-ph].
\bibitem{Lu:2022bgw} C. T. Lu, L. Wu, Y. Wu and B. Zhu,
{\it Electroweak Precision Fit and New Physics in light of $W$ Boson Mass},
[arXiv:2204.03796 [hep-ph]].
\bibitem{Heckman:2022the} J. J. Heckman,
{\it Extra $W$-Boson Mass from a D3-Brane}, [arXiv:2204.05302 [hep-ph]].
\bibitem{Rodriguez:2019mwf}M. C. Rodriguez, 
{\it The Minimal Supersymmetric Standard Model (MSSM) and General Singlet Extensions of the MSSM (GSEMSSM), a short review}, 
[arXiv:1911.13043 [hep-ph]].
\bibitem{kazakov1}D. I. Kazakov, 
\emph{Beyond the Standard Model (In search of supersymmetry)}, 
[hep-ph/0012288].
\bibitem{Chung:2003fi} D. J. H. Chung, L. L. Everett, G. L.Kane,
S. F. King, J. D. Lykken and L. T. Wang, 
{\sl Phys.Rept.}{\bf 407}, 1 (2005). 
\bibitem{gladkaza}A. V. Gladyshev and D. I. Kazakov, 
\emph{Is (Low Energy) SUSY still alive?}, [arXiv:1212.2548 [hep-ph]].

\bibitem{Rodriguez:2022gfq}M. C. Rodriguez,
\emph{The Minimal Supersymmetric Standard Model (MSSM) with $R$-Parity Violation},
[arXiv:2204.05348 [hep-ph]].

\bibitem{Montero:2007cd}J. C. Montero and V. Pleitez,
{\it Gauging U(1) symmetries and the number 
of right-handed neutrinos},
{\sl Phys. Lett.}{\bf B675}, 64, (2009); 
[arXiv:0706.0473 [hep-ph]].


\bibitem{Montero:2016qpx} J. C. Montero, V. Pleitez, M. C. Rodriguez 
and B. L. S\'anchez-Vega,
{\it Supersymmetric $U(1)_{Y^{\prime}}\otimes U(1)_{B-L}$ extension of the standard model},
{\sl Int. J. Mod. Phys.}{\bf A32}, 1750093, (2017), 
[arXiv:1609.08129 [hep-ph]].

\bibitem{Montero:2016qpxno} M. C. Rodriguez,
{\it The Supersymmetric $(B-L)$ model with three nonidentical right-handed neutrinos},
{\sl Int. J. Mod. Phys.}{\bf A32}, 2150038, (2021), 
[arXiv:1609.08129 [hep-ph]].


\bibitem{Bernal:2018aon}
N. Bernal, D. Restrepo, C. Yaguna and \'O. Zapata,
{\it Two-component dark matter and a massless neutrino in a new $B-L$ model},
{\sl Phys. Rev.}{\bf D99}, 015038, (2019); [arXiv:1808.03352 [hep-ph]].


\bibitem{Montero:2011jk} J. C. Montero and B. L. Sanchez-Vega,
{\it Neutrino masses and the scalar sector of a B-L extension of the standard model},
{\sl Phys. Rev.}{\bf D84}, 053006, (2011); [arXiv:1102.0321 [hep-ph]].
\bibitem{Sanchez-Vega:2014rka} B. L. S\'anchez-Vega, J. C. Montero 
and E. R. Schmitz,
{\it Complex Scalar DM in a B-L Model},
{\sl Phys. Rev.}{\bf D90}, 055022 (2014);  arXiv:1404.5973 [hep-ph]].


\bibitem{Pankov:2019yzr} A. A. Pankov, P. Osland, I. A. Serenkova and V. A. Bednyakov,
\emph{High-precision limits on $W–W'$ and $Z–Z'$ mixing from diboson production using the full LHC Run 2 ATLAS data set},
\emph{Eur. Phys. J.}{\bf C80}, 503 (2020).
\bibitem{Osland:2020onj} P. Osland, A. A. Pankov and I. A. Serenkova,
\emph{Updated constraints on $Z'$ and $W'$ bosons decaying into bosonic and leptonic final states using the run 2 ATLAS data},
\emph{Phys. Rev.}{\bf 103}, 053009, (2021).


\end{thebibliography}
\end{document}